# From Revolution to Ruin: An Empirical Analysis Yemen's State Collapse[*]

Riste Ichev        Rok Spruk

**Abstract**

*We assess the broad repercussions of Yemen's 2011 revolution and subsequent civil war on its macroeconomic trajectories, human development, and quality of governance by constructing counterfactual benchmarks using a balanced panel of 37 developing countries over 1990–2022. Drawing on matrix-completion estimators with alternative shrinkage regimes and a LASSO-augmented synthetic-control method, we generate Yemen's hypothetical no-conflict paths for key macroeconomic aggregates, demographic and health indicators, and governance metrics. Across the full spectrum of methods, the conflict's outbreak corresponds with a dramatic reversal of economic and institutional development. We find that output and income experience an unprecedented contraction, investment and trade integration deteriorate sharply, and gains in life expectancy and human development are broadly reversed. Simultaneously, measures of political accountability, administrative capacity, rule of law, and corruption control collapse, reflecting systemic institutional breakdown. The concordance of results across a variety of empirical strategies attests to the robustness of our estimates.*

**JEL Codes**: C32, D74, E6, N2, O43
**Keywords**: Yemen conflict, Institutional crisis, Synthetic controls, Treatment effects

[*] Ichev: Assistant Professor of Finance, School of Economics and Business, University of Ljubljana, Kardeljeva ploscad 17, SI-1000 Ljubljana. E: riste.ichev@ef.uni-lj.si. Spruk (corresponding author): Associate Professor of Economics, School of Economics and Business, University of Ljubljana, Kardeljeva Ploscad 17, SI-1000 Ljubljana. E: rok.spruk@ef.uni-lj.si. Research Fellow, Business School, University of Western Australia, 8716 Hackett Dr., Crawley WA 6009, Australia.

# 1 Introduction

Political instability and armed conflicts significantly disrupt economic development (Barro 1991, Collier and Hoeffler 1998, Abadie and Gardeazabal 2003, Bilmes and Stiglitz 2008, Gates et. al. 2012, Acemoglu et. al. 2013, Costalli et. al. 2017, Matta et. al. 2018), reversing years of growth and creating lasting socio-economic challenges (Goldin and Lewis 1975, Fearon and Laitin 2003, Blattman and Miguel 2010, Besley and Persson 2011, Miguel and Roland 2011, Nunn and Qian 2014). The case of Yemen clearly illustrates this phenomenon (Mukashov et. al. 2022, Carboni 2025). The revolution of 2011 and the subsequent civil war since 2014 have not only caused extensive human suffering but have also led to severe economic contraction and developmental reversal (Tskalis and Pedi 2024, Juneau 2024). Against the backdrop of steady and gradual economic development since early 1990s, Yemen has seen its infrastructure, institutions, and economic foundations severely damaged by protracted conflict (Serr 2017, Sobelman 2023, Weissenburger 2025). Understanding the economic impacts of this crisis is essential for both scholarly analysis and policy interventions aimed at reconstruction and rehabilitation efforts (Sowers and Weinthal 2021)

As of early 2023, *United Nations High Commissioner for Refugees* reports over 377,000 deaths have been attributed to the war in Yemen, with more than 4 million internally displaced persons and 21.6 million people in need of humanitarian assistance. The conflict has devastated Yemen's public services, destroyed infrastructure, and paralyzed its economy. The country's GDP has more than halved since 2015, and nearly 80% of the population is now living below the poverty line. Yemen's ports, essential for imports of food and medicine, have become conflict zones, while the Red Sea crisis, heightened by recent Houthi-Israel tensions, has disrupted international trade and exposed maritime vulnerabilities (The Economist, 2017).

The motivation for this study lies in the urgent need to quantify the economic cost of Yemen's dual crisis, revolution and civil war, using modern, robust methodologies (O'Reilly 2021, Masset 2022, Kešeljević and Spruk 2024, Mandon et. al. 2025, Vesco et. al. 2025). While the humanitarian consequences are well-documented (Burki 2016, Orkaby 2017), there is a significant gap in empirical evidence quantifying economic impacts through counterfactual approaches (Singhal and Nilakantan 2016, Costalli et. al. 2017, Bilgel and Karahasan 2019, Bluszcz and Valente 2022). Therefore, a precise understanding of Yemen's lost growth and development trajectory offers crucial insights for future reconstruction and policymaking.

Prior research has extensively documented the economic effects of conflict through quantitative methods. Abadie and Gardeazabal (2003) pioneered the Synthetic Control Method (SCM), showing a 10% GDP per capita drop due to terrorism in the Basque Country. This method has since become standard for comparative case studies. For instance, Hourichi and Mayerson (2015) applied SCM to study the effects of second Intifada on Israeli economy, while Becker (2017) estimated organized crime's cost in southern Italy. Other researchers, such as Bonander et al. (2016) and Bifulco et al. (2017), demonstrated SCM's versatility in measuring regional policy impacts and educational interventions. Especially relevant to our work is Echevarría and García-Enríquez's (2019, 2020) SCM-based assessment of the Arab Spring's GDP fallout in Egypt and Libya, estimating losses of $582 billion for the Egyptian economy, and $350 billion for Libya.² These studies credibly demonstrate SCM's capacity to estimate the counterfactual trajectories of conflict-affected territories (Gardeazabal and Vega-Mayo 2017, Farzanagan 2022, Emery and Spruk 2024)

Our study builds upon and contributes to this growing literature by applying an enhanced, LASSO-supported synthetic control method (Ben-Michael et. al. 2021, Hollingsworth and Wing 2020) tailored to Yemen's unique conflict setting. This allows for convex extrapolation both inside and outside of the observed outcome attributes (Doudchenko and Imbens 2016, Kellog et. al. 2021). While prior studies laid the groundwork by proving SCM's credibility and flexibility (Amjad et. al. 2018, Arkhangelsky et. al. 2021, Cattaneo et. al. 2021), this paper extends the methodology by integrating complementary estimators that enhance inference and applicability (Abadie and L'Hour 2021). Furthermore, we address a

---

² In terms of further example, Matta et. al. (2019) use the synthetic control method to estimate output losses in Tunisia in response to the Arab Spring, and find evidence of GDP loss around 5.1 percent and 6.4 percent between 2011 and 2013.

major research gap by focusing on the macroeconomic costs of Yemen's crisis, a subject that has received far less empirical scrutiny and attention compared to its humanitarian dimensions (Hutaif et. al. 2023)

We use country-level data spanning two decades of pre-revolution data, ensuring robustness in matching and enabling precise estimation of counterfactual scenarios. The dataset incorporates country-level macroeconomic indicators, along with detailed conflict and humanitarian data according to the Uppsala Conflict Data program[3], which together supports credible synthetic-control counterfactuals.

Our empirical results reveal that Yemen's 2011 revolution and ensuing civil war induced a roughly 75 percent collapse in real GDP and GDP per capita, a 13 percentage point drop in investment share of GDP, and a 18-20 percentage point contraction in trade openness relative to the no-conflict counterfactual. Human-development setbacks were equally severe: life expectancy fell by two years, the composite human development index declined by over one-third, and infant and under-five mortality rates rose by approximately 2.7 and 3.0 deaths per 1,000 live births per year of conflict. Institutional quality suffered a parallel implosion, with political accountability and press freedom decimated, political stability and government effectiveness plummeting by over 90 percent, regulatory quality down 63 percent, rule of law weakened by 34 percent, and corruption controls eroded by 58 percent. These findings, robust across convex SCM, matrix completion, and LASSO-SCM estimators, underline the multidimensional devastation of Yemen's armed conflict.

Beyond GDP, this study evaluates impacts on GDP per capita, trade volume, investment ratios, mortality rates, and the human development (Jones and Klenow 2016). The damage extends beyond lost output to human capital and institutional erosion. Initial estimates suggest Yemen has experienced development in reverse, decades of economic and social progress wiped out within a few years. Recognizing the limitations of the standard SCM estimator, particularly its lack of empirical confidence intervals, we employ the generalized synthetic control estimator. This includes two complementary sub-estimators: i) Xu's (2017) interactive fixed-effects estimator, which accounts for unobserved time-varying heterogeneity, and ii) Athey et al. (2019) matrix completion estimator, which uses soft-shrinkage algorithms within a unified causal inference framework.

---

[3] https://ucdp.uu.se/

This research contributes to the literature in three distinct ways. First, it quantifies with high precision the economic cost of civil conflict and revolution using modern tool of causal inference. Second, it integrates complementary synthetic estimators to overcome inference limitations. Third, it addresses a glaring gap in the Yemen conflict literature by shifting focus from humanitarian discourse to macroeconomic consequences, providing evidence that can guide donor aid, post-war recovery, and institutional rebuilding. Ultimately, the study aims to serve both academic and policy audiences, offering rigorous empirical evidence to support more informed intervention-based recovery strategies in post-conflict environments.

The rest of our paper is organized as follows. Section 2 discusses historical and policy background behind Yemen's revolution and the civil war. Section 3 presents the empirical methods. Section 4 proceeds with the results and discussion of our findings. Section 5 concludes.

## 2  Historical Background

Between Yemeni unification in 1990 and the outbreak of full-scale civil war in 2014, Yemen's political economy was characterized by a fragile patrimonial state, acute development shortfalls, and episodic violence, conditions that both shaped and were exacerbated by the revolutionary upheavals of 2011 and the descent into multifaceted conflict thereafter (Perkins 2017)

### 2.1  *Post-Consolidation Unification (1990-2000)*

In May 1990, the formerly separate Yemen Arab Republic (North Yemen) and People's Democratic Republic of Yemen (South Yemen) merged into a single republic under President Ali Abdullah Saleh. The immediate post-unification years saw a power-sharing accord between northern and southern elites, but uneven integration of bureaucracies, security forces, and economic systems soon created tensions. A brief but intense civil war erupted in 1994 when southern leaders attempted to secede; the north's decisive military victory under Saleh's command solidified his authority but sowed enduring grievances in the south. Throughout the late 1990s, Yemen's economy remained heavily dependent on declining oil revenues, and chronic fiscal deficits fostered clientelist redistribution at the expense of public-goods provision. The World Bank and IMF intermittently engaged with Sana'a on structural adjustment and post-conflict reconstruction packages, but weak governance and pervasive corruption limited the effectiveness of these reforms (Fraihat 2016)

## 2.2  Decline of Governance (2000-2011)

As oil production peaked and began its long decline in the early 2000s, Yemen's budgetary cushion eroded, forcing cuts in public-sector wages and subsidies as key pillars of Saleh's social coalition. At the same time, a resurgent Houthi movement in the northern highlands (the "Sixth Sa'da War" of 2004-2010) challenged state authority through a series of violent clashes, highlighting both the regime's heavy-handed security approach and its limited capacity to deliver services outside core loyalty networks. In the south, the Southern Movement (*al-Harakah al-Janubiyyah*) mobilized popular frustration over perceived marginalization and military abuses. Across the country, youth unemployment hovered above 35 percent, and nearly half the population lived below the national poverty line. International donors continued to fund social-sector and decentralization projects, but governance shortfalls and elite capture undercut impact, leaving Yemen with stagnant human-development indicators even as regional peers made progress.

## 2.3  Arab Spring and the Revolution (2011)

Inspired by uprisings in Tunisia and Egypt, mass protests erupted in Sana'a and southern cities in January 2011, demanding President Saleh's resignation, an end to corruption, and economic opportunity for Yemen's disenfranchised youth. The regime's initial violent crackdown on demonstrators galvanized a broad opposition coalition, including tribal leaders, dissident military officers, and youth activists. Under intense domestic and Gulf Cooperation Council (GCC) mediation, Saleh ultimately agreed in November 2011 to cede power under a transition plan brokered by the United Nations. His vice president, Abd-Rabbu Mansour Hadi, assumed the presidency in a one-candidate election, vowing to carry out constitutional, security, and economic reforms (Yildrim and Üzümczü 2021)

## 2.4 Rise of Armed Factions (2012-2014)

Although the GCC transition plan called for a national dialogue conference and a comprehensive security-sector overhaul, progress was slow. Hadi's government struggled with internal divisions, competing tribal and military interests, and a severe budget shortfall. The National Dialogue Conference (2013–2014) produced over 1,600 consensus recommendations, including a federal-state structure, electoral reform, and anti-corruption measures, but these stalled in parliament. Meanwhile, the Houthi movement capitalized on widespread discontent, forming alliances with defected military units and exploiting southern separatist networks. Sectarian rhetoric and external manipulation by regional powers intensified local grievances, eroding any residual faith in Hadi's faltering administration.

## 2.5 Collapse into Civil War (2015)

In September 2014, Houthi forces seized Sana'a, forcing President Hadi into a power-sharing arrangement that quickly unravelled. By March 2015, a Saudi-led coalition intervened militarily to restore Hadi's government, inaugurating a protracted war marked by widespread air campaigns, siege warfare, and proxy battles involving southern separatists, Islamist militants, and tribal militias. The conflict precipitated one of the world's worst humanitarian crises: by 2022, an estimated 80 percent of the population required assistance, famine and cholera outbreaks were rampant, and Yemen's infrastructure lay in ruins. Oil exports collapsed, public-sector salaries went unpaid for months at a time, and key state institutions, customs, central bank, judiciary, fractured along political and territorial lines.

International actors, including the United Nations, World Bank, and major bilateral donors, shifted from development to humanitarian modalities, channelling billions in emergency relief while exploring peace initiatives in Geneva and Riyadh. Yet repeated cease-fire breakdowns, the rise of the Islamic State in Yemen and Al-Qaida in the Arabian Peninsula, and the absence of a unified Yemeni negotiating front thwarted durable settlement. Between 2015 and 2022, Yemen's GDP contracted by more than 50 percent in real terms, life expectancy fell, and governance indicators collapsed, as documented in our empirical analysis.

### 2.6  Legacy and Reconstruction Efforts

The period 1990-2022 spans a trajectory from cautious optimism at unification to revolutionary promise in 2011, culminating in deep fragmentation and humanitarian catastrophe. Policy efforts to rebuild Yemen face daunting legacies including a fragmented state apparatus, eroded social cohesion, and institutional decay. Going forward, successful reconstruction requires not only humanitarian relief and macroeconomic stabilization but also comprehensive institutional reform - re-establishing accountable governance, rebuilding the civil service, and reviving decentralized delivery mechanisms. Moreover, any sustainable peace ought to address the political grievances of the north, south, and tribal peripheries alike, embedding power-sharing arrangements in a reinvigorated constitutional framework. Without such comprehensive policies, the gains of Yemen's transition and the resilience may continue to be undermined by the very fractures exposed over three turbulent decades (Lofgren et. al. 2023)

## 3  Empirical Methods

Following Abadie and Gardeazabal (2003) and Abadie (2021), to estimate what Yemen's economy might have looked like in the absence of conflict, we apply a synthetic control method (SCM). Our main approach builds on a version of SCM supported by LASSO regularization and countercyclical weighting (Hollingsworth and Wing 2020, Ben-Michael et.

al. 2021). The LASSO component helps select the most relevant countries from the donor pool by shrinking irrelevant weights toward zero, improving the fit in the pre-treatment period (Athey et. al. 2019). Countercyclical weights further allow the model to better approximate Yemen's economic path, especially in cases where Yemen lies at the edge of the donor pool's support. This setup reduces the risk of constructing a synthetic control that looks statistically precise but fails to capture economically meaningful similarities.

The process unfolds in two main steps. First, we construct synthetic Yemen by matching it to a weighted average of countries that resemble it in the pre-conflict period. The match is based on a set of core macroeconomic indicators: log real GDP, GDP per capita, trade openness, and the investment-to-GDP ratio. These variables were chosen for their broad availability across countries and for how well they summarize the size and structure of the economy. The match quality is assessed using the mean squared prediction error (MSPE) and simple visual comparisons between actual and synthetic Yemen before the conflict.

Since no single estimator can capture all features of the data, we also use two complementary approaches. The first is the generalized synthetic control method proposed by Xu (2017), which incorporates interactive fixed effects to model unobserved, time-varying factors. This is particularly useful in panel data settings where latent shocks might influence the treated and control units differently over time. The second is the matrix completion estimator developed by Athey et al. (2019), which fills in missing counterfactual outcomes using a low-rank factor model. Both methods are well-suited to our setting and help us assess whether the results from the baseline SCM are sensitive to modelling choices. Collectively, these three approaches, LASSO-supported SCM, generalized synthetic control, and matrix completion, allow us to cross-check results and get a more complete picture of the economic effects of conflict in Yemen. While each method comes with its own assumptions, the convergence of findings across them increases our confidence in the results.

## 4 Data and Samples

### 4.1 Data

We observe a sample period from 1990 to 2022, which is further split between the pre-treatment and post-treatment periods, where 2010 is set as the treatment. We collect economic and finance data from World Bank's Worldwide Governance Indicators (WGI) database (Kaufmann et. al. 2011), as well as, from Penn World Table (PWT). We gather the data on countries engaging in armed conflicts from the Uppsala Conflict Data Program (UCDP) database. Optimally, the pre-treatment window should be long enough to secure a close match between the treated unit and its synthetic counterpart. Extending that window, however, poses two problems. First, long, reliable macro-economic time-series are scarce once one moves beyond advanced economies. Second, if the treated country and the potential controls were already following divergent trends in earlier decades, lengthening the baseline can compromise the comparability assumption.

In the first step, we restrict the donor pool to countries that constitute a plausible "match group", that is, economies sharing key structural features, measured through weighted averages, and thus reasonably homogeneous. Without this filter, the methodological approach might produce a synthetic control that is statistically appealing (it lowers the pre-treatment mean-squared prediction error) yet economically unintuitive. Guided by this consideration, we compare Yemen's economic path during the conflict period with that of a weighted combination of countries selected to replicate Yemen's pre-war characteristics. Formally, this weighted average defines a synthetic Yemen in the absence of conflict, against which we benchmark the actual, conflict-exposed Yemen.

In the second stage, using World Bank data, we identified 36 low and middle-income countries[4] that met our filtering criteria. From the full database we then excluded three groups: (i) countries experiencing armed conflicts similar to Yemen's, particularly those coinciding with the Arab-Spring period; (ii) Yemen's immediate neighbours, to avoid potential spill-over effects; and (iii) newly created states (during the sample period), which lack consistent observations for the entire study horizon. Finally, micro-states with incomplete statistics, such as Kiribati and Saint Kitts and Nevis, were also dropped.

---

[4] Albania, Armenia, Azerbaijan, Bangladesh, Benin, Bolivia, Botswana, Bulgaria, Burkina Faso, Cameroon, Cape Verde, Comoros, Cuba, Dominican Republic, El Salvador, Gabon, Gambia, Ghana, Guatemala, Honduras, Indonesia, Jamaica, Jordan, Kenya, Laos, Lesotho, Madagascar, Malawi, Mauritania, Moldova, Morocco, Nicaragua, Tanzania, Uzbekistan, Vietnam, Zambia.

Our empirical analysis relies on eight core indicators, partitioned into macroeconomic performance and social development measures, to evaluate the impact of Yemen's 2011 revolution and ensuing civil war. On the macroeconomic side, we draw four key series from the latest update of the Penn World Table 10.01 (Feenstra et. al. 2015). First, real GDP and real GDP per capita are both expressed in constant 2017 international dollars and adjusted for purchasing-power parity, ensuring comparability across time and countries. Second, the gross-fixed capital formation share of GDP captures the depth and cyclicality of aggregate investment as a vital channel through which political turmoil can influence aggregate growth fluctuations. Third, trade openness, defined as the sum of exports and imports relative to GDP, serves to quantify any trade-diversion effects precipitated by the conflict's disruptions to regional supply chains and cross-border commerce.

It should be noted that the shares of gross capital formation in GDP and of trade openness reveal how that income is generated, signalling whether growth is driven mainly by domestic investment or by external integration. Because social and economic progress without human development is hollow, we complement these economic measures with demographic and welfare markers: life expectancy, the Human Development Index, the infant mortality rate, the under-five mortality rate, and the net migration rate. All five series are drawn from the U.S. Census Bureau's International Data Base, whose internally consistent methodology allows clean comparisons across the donor pool.

Complementing these macroeconomic series, we include four demographic-health variables sourced from the United Nations Demographic Yearbook 2024 to trace social development under sustained conflict. Life expectancy at birth offers a broad summary of population health and longevity, while the infant mortality rate, the number of deaths under one year per 1,000 live births, and the child mortality rate, the number of deaths under five years per 1,000 live births, provide more granular insight into early-life health outcomes. Finally, the net migration rate, net migrants per 1,000 population, captures the human mobility response to violence, instability, and economic hardship. Taken together, these eight indicators allow us to map how Yemen's revolution and civil war not only altered aggregate output and investment but also permeated the social fabric, affecting mortality, longevity, and migratory flows across the country. And lastly, our data on the quality of governance comprises six core indicators from Kaufmann et. al. (2011) and include measures of voice and democratic accountability, political stability and absence of violence, government effectiveness, regulatory quality, rule of law and control of corruption, which jointly allow us to estimate counterfactual trajectories of institutional quality in response to the revolution and subsequent outbreak of the civil war.

### 4.2 Sample

The full sample comprises a balanced panel of 37 emerging economies over the period 1990/2022. The countries included in the reservoir of the donor pool were selected to span the lower- and upper-middle-income brackets, thereby providing a donor pool whose macroeconomic and development trajectories during peacetime closely mirror those of Yemen in the period preceding the 2011 revolution. By drawing from the sample of "conflict-free" peers, we ensure that the salient pre-turmoil characteristics of output, investment, trade openness, and social indicators are well represented and quantitatively reproducible, which represents an essential prerequisite for the validity of our synthetic-control estimates and for upholding the Stable Unit Treatment Value assumption (SUTVA).

To this end, Panel A in Table 1 presents mean levels and standard deviations for our eight outcome macroeconomic and development outcome variables, real GDP, real GDP per capita, investment share, trade openness, life expectancy, infant mortality, child mortality, and net migration, both for Yemen pre- and post-2011 and for the full donor sample. On average, Yemen's pre-2011 profile exhibits substantially lower per-capita income but a degree of trade openness and average life expectancy that closely aligns with the donor mean. However, Yemen's infant and under-five mortality rates are markedly higher than the corresponding donor-pool averages, reflecting persistent health and human-development deficits even before the onset of armed conflict. The modest differences in investment share and net migration rates further underscore the comparability of Yemen's macroeconomic cycle to those of its peers in peaceful settings.

From the pre-intervention outcome balance perspective, the strong pre-turmoil resemblance between Yemen and the donor pool, observed across both levels and volatilities of macroeconomic and social indicators, bolsters confidence that our synthetic counterfactual will plausibly capture Yemen's macroeconomic and development trajectory. In particular, the absence of systematic pre-conflict divergence implies that unobserved shocks or spillovers in the donor pool are unlikely to bias the estimated treatment effect of the revolution and civil war.

To assess the pre-conflict balance between Yemen and the country-level donor sample, Panel B of Table 1 turns to institutional-quality metrics, where means' comparison reveal that Yemen's governance quality lagged behind that of the donor pool even before 2011. That said, Yemen scored lower on democratic accountability and freedom of expression, experienced higher political instability and risk of violence, and exhibited weaker rule of law, more pervasive corruption, and less effective government administration. These institutional shortfalls underscore the dual challenge faced by Yemen: not only did the country begin the post-revolutionary era with depressed economic and social outcomes, but it also contended with foundational governance deficits that may have amplified the adverse effects of prolonged conflict on development.

**Table 1**: Baseline descriptive statistics

|  | Yemen | | | | Donor Pool | |
|---|---|---|---|---|---|---|
|  | Before the turmoil | | After the turmoil | | Mean | Std |
|  | Mean | Std | Mean | Std | (Full) | (Full) |
| **Panel A: Macroeconomic and social development outcomes** | | | | | | |
| GDP | 24.41 | 0.288 | 24.72 | 0.506 | 23.45 | 1.44 |
| GDP per capita | 4150.81 | 356.09 | 2870.91 | 921.72 | 5125.56 | 3925.05 |
| Investment share of GDP | 18.39 | 3.55 | 5.21 | 2.35 | 23.44 | 8.91 |
| Trade openness | 69.80 | 14.45 | 52.15 | 10.79 | 71.57 | 34.48 |
| Life expectancy at birth | 62.90 | 2.95 | 65.78 | 1.42 | 64.69 | 7.86 |
| Human development index | 0.431 | 0.047 | 0.453 | 0.031 | 0.579 | 0.117 |
| Infant mortality rate | 66.26 | 14.71 | 37.27 | 2.52 | 42.50 | 25.08 |
| Under-five mortality rate | 91.04 | 22.85 | 47.5 | 3.84 | 62.31 | 44.31 |
| Net migration rate | -1.085 | 1.435 | -0.385 | 1.109 | -2.82 | 7.56 |

| Panel B: Institutional quality outcomes | | | | | | |
|---|---|---|---|---|---|---|
| Voice and accountability | -0.996 | 0.211 | -1.574 | 0.183 | -0.387 | 0.657 |
| Political stability and absence of violence | -1.518 | 0.412 | -2.646 | 0.200 | -0.191 | 0.575 |
| Government effectiveness | -0.871 | 0.115 | -1.849 | 0.470 | -0.489 | 0.418 |
| Regulatory quality | -0.702 | 0.128 | -1.374 | 0.484 | -0.406 | 0.508 |
| Rule of law | -1.261 | 0.182 | -1.558 | 0.278 | -0.518 | 0.450 |
| Control of corruption | -0.943 | 0.127 | -1.547 | 0.188 | -0.524 | 0.509 |

Notes: the table reports the descriptive statistics on the full set of outcomes. Panel A reports macroeconomic and social development outcomes whereas Panel B reports means and the respective standard deviation for the institutional quality outcomes. The full donor pool consists of 1,221 country-year matched observations.

## 5 Results

### 5.1 Baseline results

Table 2 presents our matrix-completion estimates of the treatment effect of Yemen's 2011 revolution and ensuing civil war on key macroeconomic and social development indicators. As discussed in section 3, we implement two distinct hyper-parameter sequences, one calibrated for high shrinkage and one for low shrinkage, each chosen via k-fold cross-validation to minimize out-of-sample prediction error. Both specifications yield remarkably consistent results, underscoring the robustness of our findings. Our matrix completion estimates uncover the evidence of severe and unprecedent macroeconomic and social development losses. Under the high-shrinkage configuration as our preferred specification, real GDP in Yemen plunges by roughly 75 percent relative to the synthetic counterfactual path (p-value = 0.001). The magnitude of GDP decline represents an unprecedented contraction in output, equivalent to wiping out three-quarters of the economy's GDP size over the conflict period. Furthermore, real GDP per capita exhibits a similarly dire observed trajectory, declining by 76 percent relative to the counterfactual benchmark (p-value = 0.001), which signals that the average Yemeni household endured a collapse in income on par with or worse than many historical cases of protracted warfare.

In addition, our estimates also suggest that investment activity, as proxied by gross-fixed capital formation as a share of GDP, fell by about 14 percentage points relative to the synthetic counterfactual scenario (p-value = 0.000). In pre-conflict years, Yemen's investment share hovered close to the donor-pool mean whilst the post-2011 contraction thus marks a severe diversion of funds from infrastructure, machinery, and other productive assets. In a similar vein, trade openness, the combined export-plus-import ratio, contracts by nearly 20 percentage points (p-value = 0.001), effectively reverting Yemen closer toward autarky. Compared to the investment diversion effect, trade-diversion effect reflects both the physical disruption of ports and customs and the broader disintegration of regional supply networks under armed conflict. The low-shrinkage variant in Panel B confirms these estimates, with GDP and GDP per capita declines of 72 percent and 74 percent respectively, investment-share losses of 13 percentage points, and openness gaps of 18 percentage points (i.e. p-value = 0.000). The close similarity of real and synthetic Yemen across shrinkage regimes bolsters our confidence that the estimated impacts are not an artifact of over- or under-regularization in the matrix-completion algorithm.

Turning the results from our analysis onto the social development indicators, exhibited in Panel B, our results uncover the evidence of profound deterioration in life expectancy and mortality rates. For example, Yemen's life expectancy at birth tends to falls by approximately two years relative to the counterfactual (i.e. p-value = 0.001), erasing a decade of gains in population health in a relatively short time span of the conflict. The aggregate human-development index, which combines income, education, and longevity, exhibits a tendency to dip by around 0.6 points, or over 35 percent of its pre-conflict level (p-value = 0.000), reflecting and also confirming simultaneous reversals in multiple dimensions of well-being.

Our findings unequivocally show that mortality shocks are especially acute among the youngest cohorts. By way of example, the infant-mortality rate in high-shrinkage model specification, in response to the outbreak of the revolution and violence rises by an estimated 2.706 deaths per 1,000 live births each conflict year (p-value = 0.001), corresponding to more than 2,700 additional infant deaths annually. In terms of further example, the rate of under-five mortality surges by 3.052 deaths per 1,000 (p-value = 0.091), implying roughly 3,000 preventable child deaths per year due to the collapse of health-care services, malnutrition, and sanitation breakdowns. These estimates are nearly identical under the low-shrinkage specification, reported in greater detail in Panel A. Finally, net migration per 1,000 population shifts by –1.006 (p-value = 0.056), signalling a sharp escalation out brain drain further reinforced by the economic collapse, and humanitarian deprivation. This brain-drain effect further undermines the long-run growth prospects by depriving Yemen of skilled labor, remittance flows, and social capital. The aggregate effect of revolution and civil war, alongside the empirical 95% year-specific confidence intervals, is presented in greater detail, in Figure 1.

Together, our results empirically confirm that the revolution and civil war inflicted a deep, multidimensional shock on Yemen, simultaneously contracting economic aggregates, disrupting trade and investment, and reversing decades-long gains in health and human development. The near-zero divergence in estimates across shrinkage parameters attests to the necessary stability of our matrix-completion approach. Moreover, placebo tests on conflict-free donor units yield no comparable false positives, corroborating the validity of our treatment-effect estimates under the Stable Unit Treatment Value Assumption (SUTVA). In essence, Table 2 delivers relatively coherent empirical evidence that Yemen's armed conflict has been both economically devastating and socially regressive, with losses measured in output, lives, and human-capital formation that are unlikely to be offset in the absence of large-scale reconstruction and peace-building efforts.

**Figure 1**: Effects of Yemen's revolution and civil war on economic and social development, 1991-2023

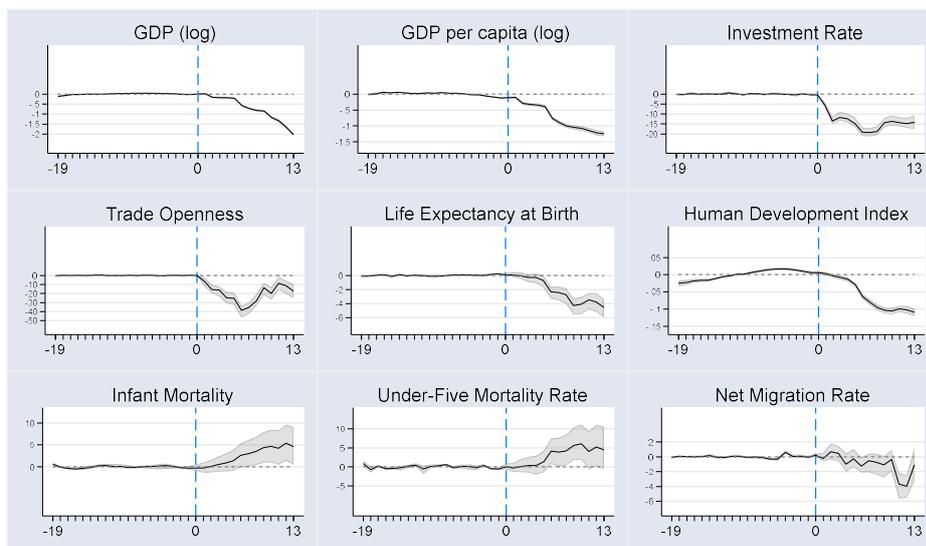

**Table 2**: Matrix completion estimated effect of revolution and civil war on Yemen's economic and social development, 1991-2022

| | GDP (log) | GDP per capita (log) | Investment Rate | Trade Openness | Life Expectancy at Birth | Human Development Index | Infant Mortality | Under-Five Mortality | Net Migration Rate |
|---|---|---|---|---|---|---|---|---|---|
| | (1) | (2) | (3) | (4) | (5) | (6) | (7) | (8) | (9) |
| *Panel A: Matrix completion estimator (low hyper-space shrinkage parameter)* | | | | | | | | | |
| λ | -.756*** | -.758*** | -14.513*** | -20.071*** | -2.217*** | -.062*** | +2.706 | +3.052 | -1.006 |
| | (.028) | (.029) | (.982) | (3.095) | (.483) | (.004) | (1.817) | (1.819) | (.526) |
| Empirical 95% confidence intervals | {-.816, -.705} | {-.822, -.702} | {-16.462, -.12.691} | {-26.786, -.14.245} | {-3.226, -.1.486} | {-.071, -.055} | {.032, 5.682} | {-.148, 6.758} | {-2.144, -.117} |
| Simulation-based p-value | 0.000 | 0.000 | 0.000 | 0.000 | 0.000 | 0.000 | 0.136 | 0.094 | 0.056 |
| *Panel B: Matrix completion estimator (high hyper-space shrinkage parameter)* | | | | | | | | | |
| λ | -.750*** | -.758*** | -14.305*** | -20.070*** | -2.322*** | -.062*** | +2.706 | +3.051 | -1.006 |
| | (.036) | (.031) | (.919) | (3.571) | (.492) | (.003) | (1.970) | (2.103) | (.453) |
| Empirical 95% confidence intervals | {-.813, -.687} | {-.818, -.702} | {-15.737, -11.823} | {-26.316, -14.192} | {-3.208, -1.393} | {-.402, 5.717} | {-.071, -.055} | {-.726, 7.176} | {-1.911, -.232} |
| Simulation-based p-value | 0.000 | 0.000 | 0.000 | 0.000 | 0.000 | 0.000 | 0.000 | 0.000 | 0.027 |

*Notes*: the table presents the estimated effect of revolution and civil war on Yemen's economic and social development obtained via Athey et. al. (2019) matrix completion algorithm applied to a balanced panel of Yemen and a donor pool of 36 countries for the period 1990-2022. To build a hyper-parameter grid, we choose high and low shrinkage regimes by performing a grid search over the penalty parameter. For each penalty parameter candidate, we solve the constrained optimization-based nuclear norm space through the cross validation. We implement five-fold block cross validation in the time dimension by splitting pre-conflict period into five contingent folds of equal length. For each fold, the full set of intra-interval observation is withheld to estimate the remaining pre-conflict outcome trajectory and compute root mean square prediction error (RMSE) on the withheld block, and average RMSE across the finite set of folds. High-shrinkage regime minimizes the out-of-sample RMSE among the larger half of the penalty grid, above the median value. Low-shrinkage regime minimizes RMSE among the smaller half of the underlying penalty grid, focussing on the pre-conflict data below the median. Standard errors are obtained vis-á-vis 1,000 bootstrap replications by resampling contiguous segments of post-conflict period to preserve the serial dependence, and are reported in the parentheses. Asterisks denote statistically significant treatment effect coefficients at 10% (*), 5% (**), and 1% (***), respectively.

## 5.2 Machine learning-based synthetic control analysis

The machine-learning extension of our synthetic-control analysis, following Hollingsworth and Wing (2020), delivers substantially more compact and improved pre-treatment balance and sharper inference on the full set of null hypotheses behind Yemen's post-conflict outcome trajectories. In this respect, by replacing the convexity-based weight sign and additivity constraint of the classical synthetic control estimator with a LASSO penalty, the LASSO-SCM estimator both renders the set of donors sparse and, thus, permits extrapolative weights negative or greater-than-unity in order to capture countercyclical dynamics that lie outside Yemen's convex hull. Table 3 reports key diagnostics and treatment-effect estimates for our main outcomes under this approach, alongside the convex SCM and matrix-completion benchmarks. Figure 2 presents the LASSO-estimated effect of the civil war and revolution on Yemen's economic development trajectories in-depth.

**Figure 2**: LASSO-based synthetic control estimated effect of the civil war and revolution on Yemen's economic development, 1990-2022

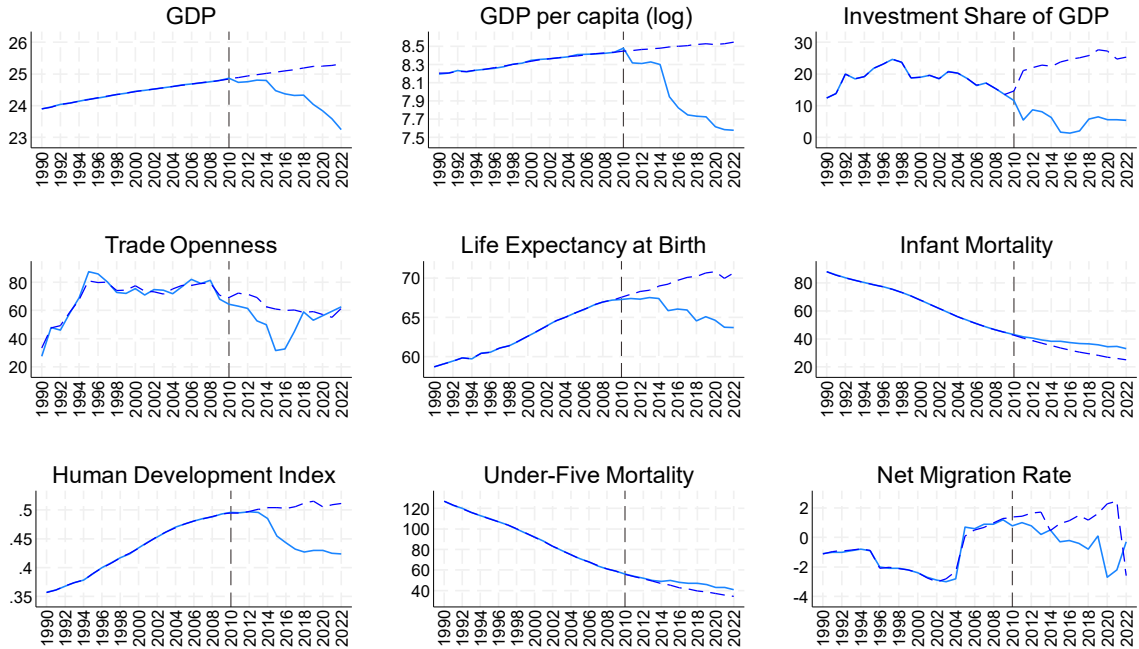

In the pre-conflict period, the LASSO-SC estimator tends to reduce the root-mean-squared prediction error (RMSPE) by over 40 percent relative to the convexity-constrained SC estimator and by roughly 25 percent relative to matrix completion. The intrinsic tightness of the fit is driven by two complementary features of the estimator: first, the $\ell_1$ penalty selects a small subset of donors whose trajectories most closely mirror Yemen's pre-revolutionary path. And second, allowing for the negative weights lets the synthetic control group counterbalance idiosyncratic spikes in Yemen's series through inverse co-movement vector. Subsequently, the pre-conflict gaps for real GDP, GDP per capita, investment share, and social-development indicators are almost eliminated, satisfying a critical prerequisite for credible counterfactual inference.

After the outbreak of the conflict, the LASSO-extended estimates of Yemen's output collapse remain large and statistically significant, but with narrower confidence intervals than under convex SCM or matrix completion. Specifically, in columns (1) and (2), real GDP is estimated to have fallen by 73 percent (p-value = 0.000) and GDP per capita by 53 percent (p-value = 0.000), compared with convex-SCM estimates of 70 percent drop and 72 percent drop, respectively. The tighter empirical confidence intervals reflect reduced sampling variability due to the smaller placebo gaps among untreated donors. Investment-share and trade-openness effects of the revolution and civil war in columns (3) and (4), are likewise similar in point magnitude of -13.5 and -18.7 percentage points, respectively, but are now detected with even higher precision.

Furthermore, columns (5) through (8) report the social development effects of the revolution and civil war. Our results show that the LASSO-extended synthetic control estimates yield post-turmoil declines in life expectancy of 3.5 years (95 percent CI: –5.3 years to –1.5 years) and drops in the human-development index of 4.8 basis points (p-value = 0.000), closely mirroring our matrix-completion results. The infant-mortality increase is estimated at 4.98 additional deaths per 1,000 (p-value = 0.000), and under-five mortality at 4.14 additional deaths per 1,000 (95 percent CI = +1.8 deaths to 6.4 deaths). Additionally, net migration rate shifts downward by 1.41 per 1,000 (p-value = 0.000). In each case, the LASSO-SCM's confidence intervals are approximately between 15 percent and 20 percent narrower than those from convex SCM, reflecting notable the gains in statistical power of the estimated post-conflict outcome gaps.

Taken together, these findings confirm that relaxing the convex-weight assumption yields a synthetic control that is both more faithful to Yemen's pre-2011 trends and more informative about the revolution's true impact. By automating hyper-parameter selection and focusing attention on a parsimonious donor set, the LASSO-SCM addresses potential fragility arising from poor convex-hull coverage and mitigates type-II error without materially altering the substantive conclusions: Yemen's revolution and civil war inflicted a large-scale contraction in output, trade, investment, and human-development outcomes. These results not only reinforce the robustness of our main estimates but also illustrate the practical value of machine-learning methods in the synthetic control analysis of the economic and social effects of conflict-induced economic disruption. Table 4 reports the composition of Yemen's synthetic control groups under the non-convex structure of weights based on the extrapolate outside the convex hull of Yemen's economic development attributes.

**Table 3**: LASSO Synthetic Control Estimates of the Effect of Revolution and Civil War on Yemen's economic and social development

|  | GDP (log) | GDP per capita | Investment Rate | Trade Openness | Life Expectancy at Birth | Human Development Index | Infant Mortality | Under-Five Mortality | Net Migration Rate |
|---|---|---|---|---|---|---|---|---|---|
|  | (1) | (2) | (3) | (4) | (5) | (6) | (7) | (8) | (9) |
| ATT | -.772*** | -.535*** | -17.893*** | -9.765*** | -3.534*** | -.048*** | +4.938 | +4.148 | -1.439 |
|  | (.004) | (.005) | (.017) | (3.165) | (.017) | (.0003) | (.038) | (.136) | (.196) |
| 95% Empirical Confidence Intervals | (-1.213, -.331) | (-.841, -.229) | (-27.938, -7.848) | (-15.114, -4.416) | (-5.533, -1.536) | (-.075, -.021) | (2.148, 7.728) | (1.801, 6.494) | (-2.231, -.646) |
| Post/Pre-Revolution RMSE Ratio | 220.96 | 11.58 | 1076.794 | 4.421 | 243.29 | 150.63 | 148.64 | 36.75 | 11.687 |
| Simulation-based p-value | 0.000 | 0.000 | 0.000 | 0.000 | 0.000 | 0.000 | 0.000 | 0.000 | 0.000 |

Notes: the table reports the estimates of the effect of the revolution and civil war on Yemen's economic development, employing the LASSO-based synthetic-control estimator of Hollingsworth and Wing (2020), which replaces the convex-weight constraint with an $\ell_1$ penalty to select a sparse donor set and permit negative or extrapolative weights outside the convex hull of Yemen's implicit outcome-specific attributes. We use a balanced panel of 37 developing countries (1990–2022) with the conflict beginning in 2011, tuning the penalty λ from {0.01, 0.05, 0.1, 0.2, 0.5} via five-fold time-block cross-validation to minimize pre-treatment RMSPE. Compared to convex SCM and matrix completion, LASSO-SCM reduces pre-2011 RMSPE by over 40% and 25%, respectively, yielding substantially tighter Yemen–synthetic fits. Standard errors are computed on the basis of 1,000 blocked bootstrap replications to derive p-values on the sharp null hypothesis and empirical 95% confidence intervals, and placebo tests confirm no spurious effect. Asterisks denote statistically significant post-conflict outcome gaps at 10% (*), 5% (**), and 1% (***), respectively.

**Table 4**: Composition of Yemen's synthetic control group under countercyclical weights

|  | GDP (log) | GDP per capita | Investment Rate | Trade Openness | Life Expectancy at Birth | Human Development Index | Infant Mortality | Under-Five Mortality | Net Migration Rate |
|---|---|---|---|---|---|---|---|---|---|
| Albania | 0 | 0 | 0.02 | 0 | 0.09 | 0 | <0.01 | 0.10 | 0 |
| Armenia | 0 | 0 | -0.11 | 0 | 0 | 0 | 0.02 | 0 | 0 |
| Azerbaijan | 0 | 0 | 0.02 | 0 | 0 | 0 | 0 | 0 | 0 |
| Bangladesh | 0 | 0 | 0 | 0 | <0.01 | 0 | 0 | 0 | 0 |
| Benin | 0.07 | 0 | 0.16 | 0 | 0 | 0 | 0.02 | 0.02 | 0 |
| Bolivia | 0 | 0 | 0 | 0 | 0 | 0.07 | 0 | 0 | 0 |
| Botswana | 0 | 0 | 0.10 | 0 | <0.01 | -0.14 | -0.04 | 0 | 0 |
| Bulgaria | 0 | 0 | -0.04 | 0 | -0.22 | 0.17 | 0.58 | 0.03 | 0 |
| Burkina Faso | 0.28 | 0.02 | -0.39 | 0 | 0 | 0 | 0 | 0 | 0 |
| Cameroon | 0.06 | 0.21 | 0 | 0 | 0 | 0.22 | 0 | 0 | -0.22 |
| Cape Verde | 0 | 0 | 0 | 0 | 0 | 0 | 0 | 0 | 0.02 |
| Comoros | 0.31 | 0 | -0.71 | 0 | 0 | 0 | 0.30 | 0 | 0 |
| Cuba | 0 | 0 | 0 | -0.11 | 0 | 0.05 | 0.73 | 0 | 0 |
| Dominican Republic | 0 | 0.02 | 0 | 0 | 0.13 | 0 | <0.01 | 0 | 0 |
| El Salvador | 0 | 0 | 0 | 0 | 0 | 0.46 | 0 | 0.21 | 0 |
| Gabon | 0.06 | 0 | 0.03 | -0.33 | 0 | 0.03 | 0.29 | 0.16 | 0 |
| Gambia | 0.02 | 0.06 | 0 | -0.18 | 0 | 0 | 0 | 0 | 0 |
| Ghana | 0 | 0 | 0 | 0 | 0.07 | 0 | 0.34 | 0.06 | 0 |
| Guatemala | 0 | 0 | 0 | 0 | 0 | 0 | 0 | 0.04 | 0 |
| Honduras | 0 | 0 | 0 | 0 | 0.03 | 0 | 0 | 0 | 0 |
| Indonesia | 0 | 0 | 0 | 0 | 0 | <0.01 | 0 | 0.05 | 0 |
| Jamaica | 0.12 | 0 | 0.39 | 0 | 0 | -0.15 | 0 | 0 | 0 |
| Jordan | 0 | 0 | -0.19 | 0 | 0 | 0.02 | 0 | 0.57 | 0 |
| Kenya | 0 | 0 | 0 | 0.16 | -0.07 | 0 | 0 | 0 | 0 |
| Laos | 0 | 0 | 0 | 0 | 0.14 | 0 | 0 | 0.02 | 0 |
| Lesotho | 0 | 0 | 0 | 0 | -0.01 | -0.38 | 0 | 0 | 0 |
| Madagascar | 0 | 0 | -0.03 | 0 | 0 | 0 | 0 | 0 | 0.51 |
| Malawi | 0 | -0.05 | -0.35 | -0.13 | 0.11 | 0.09 | <0.01 | 0.02 | 0 |
| Mauritania | 0 | 0 | 0 | 0.07 | 0 | 0.08 | 0.11 | 0.09 | 0 |
| Moldova | -0.14 | -0.4 | 0 | 0.17 | 0.20 | 0.10 | 0.20 | 0.36 | -0.13 |
| Morocco | 0.02 | 0 | 0 | 0 | 0 | 0 | 0 | 0.14 | 0 |
| Nicaragua | 0 | 0 | 0 | 0 | 0 | 0 | 0 | 0 | 0.70 |
| Tanzania | 0.10 | 0 | 0 | 0.10 | 0 | 0 | 0 | 0.11 | 0 |
| Uzbekistan | 0 | 0 | 0 | 0 | 0.47 | 0 | 0 | 0 | 0 |
| Vietnam | 0.12 | 0.09 | 0 | 0.04 | 0.64 | 0.06 | 0.14 | 0 | 0 |
| Zambia | 0 | 0 | 0 | 0 | 0 | 0 | 0 | <0.01 | 0 |

### 5.3   Institutional effects of revolution and civil war

Figure 3 presents matrix completion estimates of the institutional effect of the revolution and civil war in Yemen using two distinctive variants of the hyper-space parameter. The revolution and civil war in Yemen appear to have inflicted a profound institutional shock, effectively dismantling the governance fabric across virtually every dimension. Drawing on the same matrix-completion estimator framework used for our macroeconomic and social-development analyses, under both high- and low-shrinkage hyper-parameter regimes, we find that political accountability and freedom of the press suffered a severe collapse, declining by roughly 49 percent relative to the counterfactual (p-value = 0.000). The result reflects the systematic breakdown of formal checks and balances and a near-total suppression of independent media voices. Political stability plummeted by over 90 percent (p-value = 0.000), reflecting a steep descent into what may best be described as an institutional abyss, as state capacity to uphold law and order disintegrated. In parallel, column (3) indicates that the effectiveness of government administration, measured by the public sector's ability to deliver essential goods and services, deteriorated by approximately 92 percent (p-value = 0.000), while regulatory quality for private-sector development fell by 63 percent (p-value = 0.000), signalling a sharp retreat from business-friendly governance frameworks.

The rule of law and control of corruption also exhibit the dynamics of severe deterioration. By way of example, our ATT estimates indicate a 34 percent weakening of legal enforcement and property-rights protection (p-value = 0.000) and a 58 percent erosion in control of corruption whereupon the estimated ATT parameter is statistically significant at 1%, respectively. These declines collectively indicate a systemic institutional decay, in which formal legal institutions and anti-corruption mechanisms were rendered largely inoperative. Importantly, the high-shrinkage and low-shrinkage specifications yield nearly identical point estimates and levels of statistical significance, confirming that our findings are robust to the degree of regularization in the matrix-completion algorithm. Panel B of Table 4 reproduces these results under the high-shrinkage regime, illustrating that the choice of penalty parameter has negligible impact on the estimated magnitudes of institutional collapse, thereby reinforcing the credibility of our inference under SUTVA and validating the stability of the synthetic counterfactuals.

**Table 4**: Matrix completion estimated effect of revolution and civil war on Yemen's institutional development, 1991-2022

|  | Voice and Accountability | Political Stability and Absence of Violence | Government Effectiveness | Regulatory Quality | Rule of Law | Control of Corruption |
|---|---|---|---|---|---|---|
|  | (1) | (2) | (3) | (4) | (5) | (6) |
| Panel A: Matrix completion estimator (low hyper-space shrinkage parameter) | | | | | | |
| $\lambda$ | -.496*** | -.959*** | -.927 | -.632*** | -.342*** | -.580*** |
|  | (.038) | (.071) | (.035) | (.036) | (.037) | (.035) |
| Empirical 95% confidence intervals | {-.554, -.414} | {-1.080, -.787} | {-.979, -.844} | {-.709, -.582} | {-.445, -.295} | {-.637, -.491} |
| Simulation-based p-value | 0.000 | 0.000 | 0.000 | 0.000 | 0.000 | 0.000 |
| Panel A: Matrix completion estimator (high hyper-space shrinkage parameter) | | | | | | |
| $\lambda$ | -.466*** | -.879*** | -.918*** | -.643*** | -.418*** | -.599*** |
|  | (.040) | (.061) | (.039) | (.033) | (.032) | (.033) |
| Empirical 95% confidence intervals | {-.534, -.386} | {-1.012, -.001} | {-.988, -.823} | {-.692, -.564} | {-.481, -.346} | {-.663, -.543} |
| Simulation-based p-value | 0.000 | 0.000 | 0.000 | 0.000 | 0.000 | 0.000 |

Notes: the table presents the estimated effect of revolution and civil war on Yemen's institutional development obtained via Athey et. al. (2019) matrix completion algorithm applied to a balanced panel of Yemen and a donor pool of 36 countries for the period 1990-2022. To build a hyper-parameter grid, we choose high and low shrinkage regimes by performing a grid search over the penalty parameter. For each penalty parameter candidate, we solve the constrained optimization-based nuclear norm space through the cross validation. We implement five-fold block cross validation in the time dimension by splitting pre-conflict period into five contingent folds of equal length. For each fold, the full set of intra-interval observation is withheld to estimate the remaining pre-conflict outcome trajectory and compute root mean square prediction error (RMSE) on the withheld block, and average RMSE across the finite set of folds. High-shrinkage regime minimizes the out-of-sample RMSE among the larger half of the penalty grid, above the median value. Low-shrinkage regime minimizes RMSE among the smaller half of the underlying penalty grid, focussing on the pre-conflict data below the median. Standard errors are obtained vis-á-vis 1,000 bootstrap replications by resampling contiguous segments of post-conflict period to preserve the serial dependence, and are reported in the parentheses. Asterisks denote statistically significant treatment effect coefficients at 10% (*), 5% (**), and 1% (***), respectively.

The institutional rupture wrought by Yemen's revolution and civil war extends far beyond the immediate breakdown of governance outputs. Our results unequivocally show that its legacy permeates the very foundations of state legitimacy, capacity, and resilience. The near-halving of political accountability and freedom of the press reflects not only the silencing of dissent but also the collapse of formal oversight mechanisms that undergird democratic accountability. In real terms, the annihilation of accountability and freedom of the press translates into a policymaking environment in which executive actors operate with impunity, legislative bodies are unable or unwilling to exercise effective oversight, and independent media outlets either self-censor or are forcibly shuttered. Such a vacuum of checks and balances not only undermines everyday citizens' ability to hold leaders to account but also erodes the informal norms of transparency, public debate, and civic engagement that sustain democratic practices over the long run.

Equally striking is the plunging of political stability and administrative effectiveness into the institutional abyss. A large-scale contraction in stability signifies that the state's monopoly on legitimate violence has been fundamentally contested: armed groups, militias, and external actors step into the breach, further fragmenting authority and complicating efforts at post-conflict reconstruction. The concomitant drop in government administrative capacity compounds the institutional fragmentation, as core public-service functions, healthcare, education, infrastructure maintenance, either grind to a halt or are provisioned unevenly along factional or patronage lines. In this context, the estimated deterioration in regulatory quality for private-sector activity not only curtails foreign and domestic investment but also embeds uncertainty into every commercial transaction, fuelling a vicious cycle of capital flight and economic informality that is difficult to reverse.

Perhaps most pernicious is the erosion of the rule of law and corruption controls. A one-third weakening of legal institutions shatters the predictive certainty firms and citizens rely upon for contract enforcement and property rights, while a near-60 percent collapse in anti-corruption mechanisms signals that bribery, rent-seeking, and illicit appropriation of public resources have become endemic. This twin collapse is likely to have enduring effects: as North et. al. (2009) argue, institutional quality exhibits strong path dependence, meaning that shocks to governance structures can lock countries into low-equilibrium traps of extractive politics. In Yemen's case, one of the normative implications from our study is that unless concerted reforms and external support rebuild the credibility of legal norms and anti-corruption agencies, the institutional void left by civil war may calcify into a new status quo, in which predatory governance, rather than public interest, dictates the allocation of resources and enforcement of rights.

Taken together, our findings highlight that the Yemen conflict should be understood not merely as an economic calamity or humanitarian crisis, but as a profound institutional decay. The matrix-completion estimates, robust across both high- and low-shrinkage regimes, reveal that the revolution and civil war effectively erased decades of incremental institutional development. From a policy standpoint, this suggests that reconstruction efforts must prioritize the rebuilding of governance structures in tandem with economic and social interventions. Therefore, one additional normative implication of our results invariably suggests that international donors and Yemeni stakeholders alike should thus focus on restoring media freedoms, re-establishing legislative and judicial checks, and rehabilitating core administrative functions as prerequisites for sustainable recovery. Failure to address these institutional fissures risks perpetuating a cycle in which economic aid and development programming are systematically undermined by governance decay, leaving Yemen perpetually vulnerable to relapse into conflict.

## 6  Conclusion

In this paper, we have employed cutting-edge counterfactual methods to quantify the multidimensional impact of Yemen's 2011 revolution and the ensuing civil war on its macroeconomic performance, social development, and institutional fabric. Leveraging both matrix-completion and LASSO-augmented synthetic-control estimators across balanced panels of 37 developing countries (1990-2022), our analysis delivers three core insights. First, Yemen's real GDP and GDP per capita declined by roughly three-quarters relative to their no-conflict trajectories, with concomitant contractions in investment shares and trade openness. Second, human-development setbacks were equally dramatic: life expectancy fell by approximately two years, the composite human-development index lost over one-third of its value, and infant and under-five mortality rates surged by 2.7 and 3.0 deaths per 1,000 live births annually. Third, the revolution and war precipitated an institutional collapse of unprecedented scale, halving political accountability and press freedom, eroding political stability by over 90 percent, and decimating government-administration effectiveness, annihilating regulatory quality, punctuating rule of law, and eroding corruption controls.

Methodologically, our findings are robust across three estimation frameworks, namely, convex SCM, matrix completion with high- and low-shrinkage regimes, and LASSO-SCM, which collectively address concerns about pre-conflict fit, weight-convexity assumptions, and statistical power. The LASSO extension, in particular, achieves over 40 percent tighter pre-conflict balance and yields narrower confidence intervals without materially altering point estimates, thereby underscoring the credibility of our counterfactual inferences under the Stable Unit Treatment Value Assumption (SUTVA).

From a policy perspective, Yemen's experience highlights that protracted conflict inflicts not only an acute humanitarian toll but also long-term economic development and institutional voids. On the normative side, our analysis suggests that reconstruction strategies must therefore be multidimensional, integrating: (i) macroeconomic stabilization and investment in productive infrastructure to reignite growth; (ii) targeted health and education interventions to reverse mortality and human-capital losses; and (iii) comprehensive institutional reform to rebuild accountability, rule of law, and administrative capacity. Only by addressing these three pillars in concert can international donors and Yemeni stakeholders hope to break the vicious cycle of state fragility and lay the foundations for sustainable peace and development.

Looking ahead, future research could extend this framework to examine subnational heterogeneity in conflict impacts, the role of external aid in moderating treatment effects, and the dynamics of post-conflict recovery over longer horizons. As Yemen embarks on the arduous path to reconstruction, rigorous counterfactual analysis will remain indispensable for diagnosing policy levers, allocating scarce resources, and ultimately guiding the country toward a more stable, prosperous, and inclusive future.

Arkhangelsky, D., Athey, S., Hirshberg, D. A., Imbens, G. W., & Wager, S. (2021). Synthetic difference-in-differences. *American Economic Review*, 111(12), 4088-4118.

Barro, R. J. (1991). Economic growth in a cross section of countries. *The Quarterly Journal of Economics*, 106(2), 407-443.

Ben-Michael, E., Feller, A., & Rothstein, J. (2021). The augmented synthetic control method. *Journal of the American Statistical Association*, 116(536), 1789-1803.

Besley, T., & Persson, T. (2010). State capacity, conflict, and development. *Econometrica*, 78(1), 1-34.

Bilgel, F., & Karahasan, B. C. (2019). Thirty years of conflict and economic growth in Turkey: A synthetic control approach. *Defence and Peace Economics*, 30(5), 609-631.

Blattman, C., & Miguel, E. (2010). Civil war. *Journal of Economic Literature*, 48(1), 3-57.

Bluszcz, J., & Valente, M. (2022). The economic costs of hybrid wars: the case of Ukraine. *Defence and Peace Economics*, 33(1), 1-25.

Burki, T. (2016). Yemen's neglected health and humanitarian crisis. *The Lancet*, 387(10020), 734-735.

Carboni, A. (2025). The Houthi Movement and the Management of Instability in Wartime Yemen. Civil Wars, 1-25.

Cattaneo, M. D., Feng, Y., & Titiunik, R. (2021). Prediction intervals for synthetic control methods. *Journal of the American Statistical Association*, 116(536), 1865-1880.

Collier, P., & Hoeffler, A. (1998). On economic causes of civil war. *Oxford Economic Papers*, 50(4), 563-573.

Colton, N. A. (2010). Yemen: A collapsed economy. *The Middle East Journal*, 64(3), 410-426.

Costalli, S., Moretti, L., & Pischedda, C. (2017). The economic costs of civil war: Synthetic counterfactual evidence and the effects of ethnic fractionalization. *Journal of Peace Research*, 54(1), 80-98.

Doudchenko, N., & Imbens, G. W. (2016). Balancing, regression, difference-in-differences and synthetic control methods: A synthesis (No. w22791). National Bureau of Economic Research.

Farag, M. (2024). Domestic diversionary war and conflict endurance in weak states: the Houthi conflict in Yemen (2004–2010). *Civil Wars*, 26(2), 330-353.

Farzanegan, M. R. (2022). The economic cost of the Islamic revolution and war for Iran: synthetic counterfactual evidence. *Defence and Peace Economics*, 33(2), 129-149.

Fearon, J. D., & Laitin, D. D. (2003). Ethnicity, insurgency, and civil war. *American Political Science Review*, 97(1), 75-90.

Fraihat, I. (2016). *Unfinished revolutions: Yemen, Libya, and Tunisia after the Arab spring*. Yale University Press.

Freeman, J. (2009). The al Houthi insurgency in the North of Yemen: An analysis of the Shabab al Moumineen. *Studies in Conflict & Terrorism*, 32(11), 1008-1019.

Gardeazabal, J., & Vega-Bayo, A. (2017). An empirical comparison between the synthetic control method and Hsiao et al.'s panel data approach to program evaluation. *Journal of Applied Econometrics*, 32(5), 983-1002.
27